\documentclass{appolb}
\usepackage{graphicx}

\newcommand{\Rerm}{\mathop{\rm Re}}
\newcommand{\Imrm}{\mathop{\rm Im}}

\begin{document}
\title{Lifetime and confinement of a quasi-gluon
\thanks{Presented at Excited QCD 2022 - Giardini Naxos}
}

\author{Fabio Siringo  and  Giorgio Comitini
\address{Dipartimento di Fisica e Astronomia dell'Universit\`a di Catania,\\ INFN Sezione di Catania, Via S.Sofia 64, I-95123 Catania, Italy} }

\maketitle
\begin{abstract}
The existence of genuine complex conjugated poles in the gluon propagator is discussed and related to confinement,
string tension and condensates. The existence of the anomalous poles leads to an untrivial analytic continuation from
Euclidean to Minkowski space, where the pole part of the propagator is related to the spectrum of excited quasigluons.
\end{abstract}
  
\section{Introduction}

There is some evidence, confirmed by very different approaches\cite{stingl,RGZ,sir2016,binosi,kondo18}, that the gluon propagator might have
anomalous complex conjugated poles.
Ab initio calculations, based on a screened perturbative expansion\cite{sir2016},
confirmed the existence of the complex poles and also gave a quantitative prediction for their location,
by an optimized one-loop approximation\cite{xigauge} which relies on the gauge invariance of the poles and
provides an excellent agreement with the lattice data\cite{duarte} in the Euclidean space.

If the poles were genuine, the existence of a complex mass would lead to a dynamical
mechanism for the gluon confinement, with a strong damping rate and a finite lifetime,
$c\, t\sim 10^{-15}$ m,  which would cancel the gluon from the asymptotic states\cite{damp}.
However, the dynamical description of a {\it quasigluon} in Minkowski space would
be plagued by the problem of analytic continuation since the existence of the anomalous poles
invalidates\cite{sircom2022}  the usual Wick rotation which is used for connecting the Euclidean and Minkowki
descriptions of a  field theory.

Thus, it is crucial to understand if the poles might have a  genuine physical meaning in the first place. Then, an unconventional mechanism must be deviced for connecting 
the Euclidean and Minkowski versions of the the theory\cite{sircom2022}.  We are going to discuss such points in more detail.

\section{Are the complex poles genuine?}

While the poles are required to be gauge-parameter-independent, because of Nielsen identities\cite{nielsen}, 
the whole
principal part of the gluon propagator seems to be gauge invariant, i.e. the phase of the complex residues
has been found to be invariant by the screened expansion at one loop. Actually, from first principles, the screened expansion was optimized by enforcing the gauge invariance of the whole principal part\cite{xigauge}. The resulting gluon
propagator, without any free parameter, turns out to be in excellent agreement with the lattice data in the Euclidean
space, where the principal part $\Delta_E$, i.e. its pole part,  reads
\begin{equation}
\Delta_E(p^2)=\frac{R}{p^2+M^2}+\frac{R^\star}{p^2+(M^2)^\star} 
\label{DE}
\end{equation}
with a quantitative prediction for the poles $M=0.581+0.375\> i$ GeV and for the phase of the residues, 
which is given by the ratio $\frac{\Imrm\{R\} } {\Rerm\{R\} }=3.132$.
The same principal part provides a very good approximation for the whole propagator in the IR, up to a very small
correction\cite{dispersion,xigauge}. Let us explore some physical consequences of the anomalous poles.

\subsection{String tension (short distance limit)}

In the short-distance limit, the static quark potential can be approximated by its tree-level contribution
\begin{equation}
V(r)\approx-C_F (4\pi\alpha_s) \int\frac{{\rm d}^3{\bf k}}{(2\pi)^3} \>\Delta({\bf k}^2) e^{i{\bf k\cdot r}}
=- \frac{\alpha_s C_F}{\pi\, i\, r}\, 
\int_{-\infty}^{+\infty} k{\rm d}k \>\Delta_E( k^2) e^{i kr}
\end{equation}
then inserting the principal part, Eq.(\ref{DE}), the larger contribution is given by
\begin{equation}
V(r)=-C_F \frac{\alpha_s}{r}\left[R\exp(-Mr)+R^\star\exp(-M^\star r)\right].
\end{equation}
Expanding in powers of $r$, up to an irrelevant additive constant
\begin{equation}
V(r)\approx C_F \left(2\Rerm\{R\}\right)\alpha_s\left[-\frac{1}{r}-\frac{\Rerm\{RM^2\}}{2\Rerm\{R\}}\> r+\dots\right]
\approx {\rm const.}\times \left( -\frac{1}{r}+ kr\right)
\end{equation}
where, using the above values of $M$ and $R$, as predicted by the screened expansion,
\begin{equation}
k=\frac{-\Rerm\{RM^2\}}{2\Rerm\{R\}}=0.584\>{\rm GeV}^2>0
\label{kappa}
\end{equation}
which gives a reasonable (gauge invariant) prediction for the string tension 
\begin{equation}
\sigma=\frac{4}{3}\alpha_s\, k \approx 0.2\>{\rm GeV}^2\quad if \quad \alpha_s\approx 0.3.
\end{equation}
For later reference, we observe that expanding $\Delta_E$
in powers of $1/p^2$
\begin{equation}
\Delta_E(p^2)=(2\Rerm\{R\})\left[\frac{1}{p^2}+\frac{2k}{p^4}+{\cal O}(1/p^6)\right].
\label{p4}
\end{equation}
In Eq.(\ref{kappa}), the sign of the coefficient $k$ is correct only if  $R$ and $M$ 
are complex and their phases satisfy the condition
$\tan\theta\,\tan\phi>1$. For instance, a real pole would give the wrong sign for the string tension and for
the $1/p^4$ term, giving rise to an ordinary Yukawa potential. Thus the existence of the complex poles seems to be related
to the confining nature of the static potential.

\subsection{Condensates and OPE}
As discussed in Ref.\cite{boucaud}, in the Landau gauge, by OPE the gluon propagator can be written as
\begin{equation}
\Delta_E(p^2)=\Delta_0(p^2)+\frac{N_c g^2}{4(N_c^2-1)}\frac{\langle A^2\rangle}{p^4}+{\cal O}(1/p^6)
\end{equation}
where $\Delta_0$ is the standard perturbative result and the $1/p^4$ term is provided by OPE and related to the
existence of a dimension two gluon condensate $\langle A^2\rangle$. The resulting expression provides a very
good fit of the lattice data in the  2-10 GeV  window\cite{boucaud}. As shown in Eq.(\ref{p4}), 
the principal part of the gluon already contains the $1/p^4$ correction with the correct sign:
taking $\Delta_0(p^2)\approx Z/p^2$  for  the {\it perturbative} propagator
\begin{equation}
\Delta_E(p^2)\approx Z\left[\frac{1}{p^2}+\frac{N_c g^2}{4Z(N_c^2-1)}\frac{\langle A^2\rangle}{p^4}\right]
\end{equation}
which has the same form of Eq.(\ref{p4}).
Then, up to an irrelevant renormalization factor
\begin{equation}
k=\frac{N_c }{8Z(N_c^2-1)} \left[g^2\langle A^2\rangle\right]=\frac{-\Rerm\{RM^2\}}{2\Rerm\{R\}}=0.584\>{\rm GeV}^2>0.
\end{equation}
Again, we see that the phases of $R$ and $M$ are essential for predicting the correct condensates and string tension,
with the correct sign.

\section{Analytic continuation} 

Assuming that the complex poles are genuine and related to gauge invariant physical observables, then the standard
K\"all\'en-Lehmann spectral representation is not valid\cite{dispersion,kondo18} and a straight analytic continuation
from Euclidean to Minkowski space would be obstructed by the existence of the anomalous pole. We must generalize
the spectral representation and introduce opposite rotations for the pole parts of the propagator.

\subsection{Generalized K\"all\'en-Lehmann representation}
From first principles, for $t>0$, the gluon propagator can be written in Minkowski space as
\begin{equation}
i\,\Delta^{\mu\nu}({\bf x},t)=\langle 0\vert A^\mu(0)e^{i{\bf P }\cdot {\bf x}}\,e^{-i{\hat H}{ t}}\, 
A^\nu(0)\vert 0\rangle=\sum_n \rho^{\mu\nu}_n\> e^{i{\bf p}_n\cdot {\bf x}}\> e^{-iE_nt}
\end{equation}
and complex poles can only arise from complex energy eigenvalues\cite{sircom2022},
$E_n=\pm(\omega_n\pm i\gamma_n)$, with $E_n^2=M^2, {M^\star}^2$. 
Adding the twin part, for $t<0$, the Fourier transform reads
\begin{equation}
i\Delta({\bf p},p_0)=\sum_n (2\pi)^3\delta^3({\bf p}-{\bf p}_n)\>\rho_n(p) \int_0^\infty 
\left[e^{ip_0t}+e^{-ip_0t}\right]\>e^{-iE_n t} {\rm d}t
\end{equation}
which can only be finite if we assume a {\it ``convergence principle''}
requiring  $E_n=-\omega-i\gamma$, $E_n^\prime=\omega-i\gamma=-E_n^\star$.
Then the integral is finite (as it must be) and the propagator reads
\begin{equation}
\Delta(p)=\sum_n \frac{(2\pi)^3 \left[2\rho_n(p)\,E_n\right]\delta^3({\bf p}-{\bf p_n})}
{p^2-M_n^{2}}
-\sum_n \frac{(2\pi)^3 \left[2\rho_n(p)\,E_n^*\right] \delta^3({\bf p}-{\bf p_n})}
{p^2-M_n^{*\,2}}
\end{equation}
yielding, by Lorentz invariance, for a single complex mass shell, a Minkowskian propagator
\begin{equation}
\Delta(p)=\frac{R}{p^2-M^2}-\frac{R^*}{p^2-M^{2\,*}}. 
\end{equation}
We observe the existence of the anomalous pole, in the first term, at
 $E=\pm\sqrt{{\bf p}^2+M^2}=\pm(\omega+i\gamma)$. The first pole part also has the wrong sign compared to
the straight analytic continuation of the Euclidean principal part of Eq.(\ref{DE}); while the regular pole part,
at $E^\prime=\pm E^\star=\pm\sqrt{{\bf p}^2+{M^\star}^2}=\pm(\omega-i\gamma)$ has the correct sign. Thus, the 
Minkowskian propagator is not the straight analytic continuation of the Euclidean pole part.

%uncomment the following lines to place a figure
\begin{figure}[htb]
\centerline{%
\includegraphics[width=12.5cm]{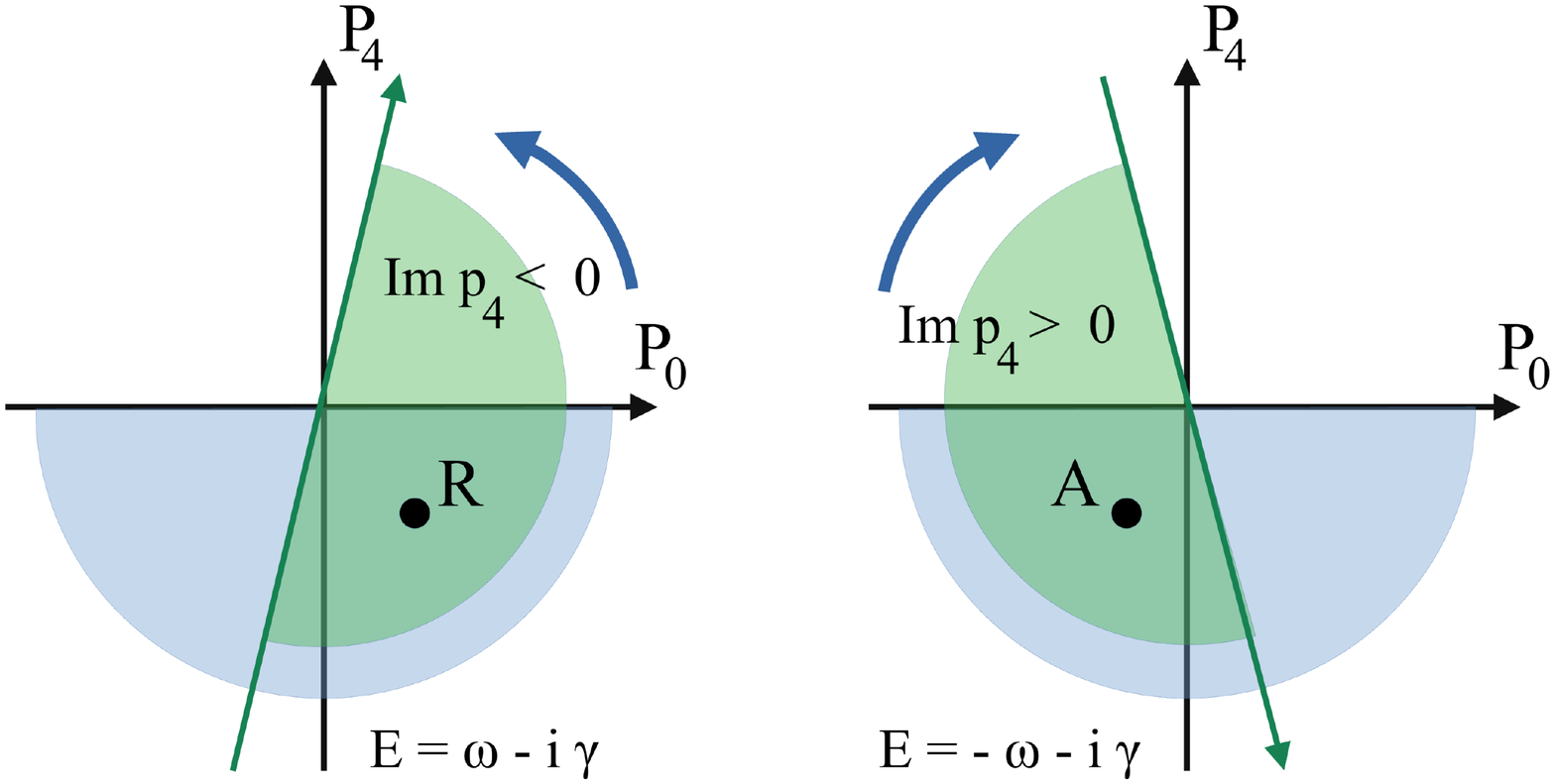}}
\caption{Clockwise and anti-clockwise Wick rotation. The anomalous pole (A) and the regular one (R) require
a clockwise and anti-clockwise Wick rotation, respectively. The shaded areas are the contours which must be chosen for $t>0$ in the Fourier transform, according to Jordan lemma.}
\label{Fig:F1}
\end{figure}

\subsection{clockwise and anti-clockwise rotation}

The opposite sign of the anomalous part can be understood by the same {\it convergence principle},
requiring a finite outcome for any physical observable or any physical quantity, like the gluon propagator, which
seems to be strongly related to physical observables, as discussed in the previous section.
Actually, when going to the Euclidean space, the usual Wick rotation cannot be used for the anomalous part because
the anomalous poles are in the wrong sectors of the complex plane and would be crossed by the usual rotation.
As shown in Fig.~1, the anomalous part of the Minkowskian propagator can only be continued to the Euclidean space
by a clock-wise rotation, which by Jordan's lemma yields the correct (finite) analytic continuation provided that a
{\it negative} imaginary time $\tau<0$ is associated to a {\it positive} real time $t>0$. In more detail, taking
$p_0=ip_4$ and $t=-i\tau$, the direct-space propagator is defined, as a function of time, 
by a Fourier transform with an exponential which reads
\begin{equation}
e^{\displaystyle{ip_0\,t}}=e^{\displaystyle{ip_4\,\tau}}. 
\end{equation}
In the transform, the contour integral is finite if and only if  $t>0$ when $\Imrm{p_0}<0$ and 
$\tau>0$ when $\Imrm{p_4}<0$. Thus, as shown in the figure and discussed in Ref.\cite{sircom2022}, 
the clockwise rotation leads to the opposite imaginary-time ordering for the anomalous part. Moreover, the reversed integration from $+\infty$ to $-\infty$, along the $p_4$
axis, leads to a change of sign for the anomalous part. Thus, the untrivial analytic continuation of the Minkowskian
propagator yields the Euclidean pole part in Eq.(\ref{DE}), with the correct sign.

\section{Conclusions}

In favour of the genuine nature of the poles, we reported their gauge invariance and their physical role in determining
the short-range string tension, condensates and, of course, the dynamical mass and damping of a gluon.
Thus, unless we accept that the real-time propagator does not even exist\cite{kondo21},
the analytic continuation of the gluon propagator must be deeply revised\cite{sircom2022}, leading to an effective Minkowskian propagator which is not given by
the trivial analytic continuation of the Euclidean function.  As shown in \cite{sircom2022}, 
the resulting Minkowskian propagator would be imaginary and defines
an added spectral density which generalizes the K\"all\'en-Lehmann representation and
might improve the reconstruction of the propagator by spectral methods.

\end{document}